\documentclass[usenatbib]{mnras}

\DeclareRobustCommand{\VAN}[3]{#2}
\let\VANthebibliography\thebibliography
\def\thebibliography{\DeclareRobustCommand{\VAN}[3]{##3}\VANthebibliography}

\usepackage{graphicx}	
\usepackage{amsmath}	

\newcommand{\im}{\imath}
\newcommand{\Var}{\mathrm{Var}}
\newcommand{\twodots}{\mathinner {\ldotp \ldotp}}

\title[Global CD/EoR Detection with Beamforming \& Beyond]{Global CD/EoR Signal Detection with a Dense Digital Beamforming Array and Beyond}

\author[J.-H. Gu et al.]{
Junhua Gu$^{1}$\thanks{E-mail: jhgu@nao.cas.cn}, Jingying Wang$^{2}$\thanks{E-mail: jywang@shao.ac.cn} , Huanyuan Shan$^{2,3}$,
Qian Zheng$^{2}$, Quan Guo$^{2}$, Yan Huang$^{1}$,  \newauthor Kuanjun Li$^{1}$,  Tianyang Liu$^{2,3}$ \\
$^{1}$National Astronomical Observatories, Chinese Academy of Sciences, 20A Datun Road, Beijing 100101, China\\
$^{2}$Shanghai Astronomical Observatory, Chinese Academy of Sciences, 80 Nandan Road, Shanghai 200030, China\\
$^{3}$School of Astronomy and Space Science, University of Chinese Academy of Sciences, 19A Yuquan Road, Beijing 100049, China
}

\date{Accepted XXX. Received YYY; in original form ZZZ}

\pubyear{2022}

\begin{document}
\label{firstpage}
\pagerange{\pageref{firstpage}--\pageref{lastpage}}
\maketitle

\begin{abstract}
The global neutral hydrogen 21 cm signal extracted from the all-sky averaged radio spectra is one of the signatures of the Cosmic Dawn and Epoch of Reionization (CD/EoR). 
The frequency-dependency of antenna beam patterns coupled with the strong foreground emission could introduce artificial spectral structures and cause false detection.
A digital beamforming array could be potentially employed to form achromatic station beam patterns to overcome this problem.
In this work, we discuss the method of forming achromatic beam patterns with a dense regular beamforming array to detect the global CD/EoR signal, covering topics including the array configuration, antenna weight optimization, and error estimation.
We also show that based on the equivalence between a beamforming array and an interferometer, most antennas in the array can be removed by canceling redundant baselines.
We present an example array design, optimize the antenna weights, and show the final array configuration by canceling redundant baselines.
The performance of the example array is evaluated based on a simulation, which provides a positive indication towards the feasibility of detecting the CD/EoR signal using a dense digital beamforming array.

\end{abstract}




\section{Introduction}
The Cosmic Dawn and Epoch of Reionization (CD/EoR) are early evolving stages of the universe predicted by today's cosmology theory. The neutral baryon matter that formed in the recombination epoch was ionized again during CD/EoR, which is expected to be initiated by the first stars \citep[e.g.,][]{2006PhR...433..181F}.
Meanwhile, the lacking of direct observational evidence makes CD/EoR a missing link in the evolving process of the universe. This topic has become one of the frontiers of modern observational cosmology and radio astronomy.
The central concerns in CD/EoR detection include a series of questions: the details of the first stars, how the neutral hydrogen was ionized, how the halo accretion process modulated the ionization process, etc.
The redshift range of the CD/EoR is expected to be about $6-27$ \citep[e.g.,][]{2010ARA&A..48..127M,1997ApJ...475..429M}, during which
the 21cm emission line produced by neutral atomic hydrogen is the essential tracer and can be detected in the meter-wave band
due to the redshift effect.

There are mainly three paradigms of detection methods for observing the redshifted neutral hydrogen 21cm line as a tracer to the CD/EoR: (1) global averaged signal detection, (2) power spectra measuring, and (3) direct imaging observation.
Among these three methods, detecting the global averaged 21cm signal requires the least observation time.
Several experiments aiming to detect the sky-averaged CD/EoR neutral hydrogen 21 cm signal are being performed, including LEDA \citep[the radiometer system][]{2018MNRAS.478.4193P, 2016MNRAS.461.2847B, 2021MNRAS.505.1575S}, SARAS \citep[][]{2013ExA....36..319P}, BIGHORNS \citep[][]{2015PASA...32....4S}, EDGES \citep[][]{2016MNRAS.455.3890M}, SCI-HI \citep[][]{2014ApJ...782L...9V}, and REACH \citep[][]{2022JAI....1150001C}.

By now, some detection results have been obtained with the EDGES experiment.
With the dipole-like antenna used in the EDGES experiment, an absorption feature (i.e., the ``dip'' feature) of $\delta T_{21}=-500_{-500}^{+200}$ mK was observed \citep{2018Natur.555...67B}.
This value is much deeper than the prediction of standard cosmology models \citep[e.g., ][]{2021MNRAS.506.5479R}, and different and/or extra mechanisms are required to explain their observational results.
Various of models have been proposed \citep[e.g., ][]{2018Natur.555...71B, 2018PhRvL.121a1101F, 2018MNRAS.480L..85H}.
On the other hand, \cite{2022NatAs...6..607S}, based on the data acquired with the SARAS 3 radiometer \citep[][]{2021arXiv210401756N}, argues that the extra deep dip found by \cite{2018Natur.555...67B} is not evidence for new astrophysics or non-standard cosmology.
More detailed cross-checks are required regarding global CD/EoR signal detection.

The frequency-dependency of the antenna beam pattern is one of the key systematics issues. As we have pointed out in our previous work \cite{2020MNRAS.492.4080G} and also been mentioned in some other works \citep[e.g.,][]{2014MNRAS.437.1056V,2015ApJ...799...90B, 2021MNRAS.506.2041A,2022MNRAS.509.4679A, 2022MNRAS.515.1580S}, the frequency-dependency of the antenna beam pattern brings extra spectral fluctuations or structures that can significantly bias the global CD/EoR signal detection.
The artifacts caused by the frequency-dependent beam are not only determined by the beam pattern itself but also by the brightness temperature angular distribution of the sky.
Theoretically, there are at least two methods to solve this problem. One is to mathematically model the antenna beam and the sky brightness temperature angular distribution and disentangle the instrument effect from the measured data \citep[]{2021MNRAS.506.2041A,2022MNRAS.509.4679A}; another is to build an instrument, the beam pattern of which is approximately frequency independent in the interested frequency range.
Most global CD/EoR detection experiments based on single antennas require the antenna to be optimized for broadband achromatic beam patterns.

Designing a single broadband antenna with frequency-independent beam patterns is challenging.
On the other hand, with digital beamforming (DBF) technology, it is possible to make the beam pattern of an array of (usually identical) antennas independent of frequency.
This idea was proposed by \cite{2020JAI.....950008D} and later improved in \cite{2021JAI....1050015D}.
The authors described the method of using the beamforming mode of the second station of the Long Wavelength Array \citep[][]{2017MNRAS.469.4537D} to generate frequency-independent beam patterns to detect the global CD/EoR signal.
A recent end-to-end simulation work by \cite{2022arXiv221004693P} in the background of the SKA Engineering Development Array v2 (EDA2) reveals that an array with $10^5$ antennas may form beam patterns with sidelobe level $\sim -50$ dB, so that can be used as an alternative method to measure the global 21 cm signal.

Another series of similar but different methods of measuring the global CD/EoR signal -- using a short-spacing interferometer -- has been suggested
\citep[e.g.,] []{2015ApJ...815...88S,2015ApJ...809...18P, 2020MNRAS.499...52M}.
This method utilizes the response of short baselines to the monopole component of the sky brightness temperature distribution.

Though DBF arrays and interferometers are regarded as different types of instruments, they share a common basic mathematical theory.
For example, \cite{ruigrok2017cross} pointed out that the auto-correlation of the output of a DBF array can be equivalently replaced by summing the weighted cross-correlation of the outputs of the single antennas that composes the DBF array.
Under this treatment, the DBF data acquisition can be performed with a standard correlator.
An array with an optimized configuration can form a beam pattern to meet the requirement of global CD/EoR signal detection. Removing redundant baselines can reduce the number of data acquisition channels to save costs.
And the number of required antennas can be reduced to several tens.

In this work, we show that a dense regular digital beamforming array, with most of its antennas ($\sim80\%$) removed following a particular procedure, can be used to form frequency-independent station beams to detect the global CD/EoR signal.

This paper is organized as follows:

Section \ref{sec_dbf_eor} describes the design of the digital beamforming array.

Section \ref{sec_dbf2corr} explains how the number of antennas (so that data acquisition channels) can be significantly reduced based on the mathematical equivalence between the DBF array and the interferometer array.

Section \ref{sec_example} gives an example design and demonstrates the effect of cutting down redundant baselines.

Section \ref{sec_discussion} presents some aspects that require further study and lists some engineering challenges that must be addressed.

Section \ref{sec_conclusion} presents our conclusions.

In the appendix, we present the deduction of the statistic error estimation.

In the following sections, especially when describing the array configuration, different indexing systems of antenna/baseline are used for different purposes, so we list the indexing-related symbols here beforehand:
\begin{description}
	\item[$s$:] the size of the regular square grid that is composed of at most $s^2$ antennas, and $s$ is chosen to be an odd number;
	\item[$(p,q)$:] antenna index on a regular square grid, i.e., the grid index, and $p=-(s-1)/2 \twodots(s-1)/2$, $q=-(s-1)/2\twodots(s-1)/2$;
	\item[$i$ and $j$:] linear antennas indices, which can be defined as $i(\mathrm{or~}j)=[p+(s-1)/2]s+q+(s-1)/2$ for an antenna with the grid index of $(p,q)$; note that the linear indices do not have to be continuous, and removing an antenna does not affect the linear indices of others;
	\item[$k$:] Redundant baselines, a group of equivalent baselines shares the same $k$ index, and $k=0$ denotes zero-length baselines, corresponding to auto-correlations.
\end{description}
\section{Designing an Aperture Array for Global CD/EoR Signal Detection}
\label{sec_dbf_eor}

\subsection{The Principle of Digital Beamforming}
Although DBF is a mature technology, we still give a brief review in the context of global CD/EoR signal detection in this section.
\label{ssec_dbf_principle}
A beamforming array or so-called \textit{aperture array} is composed of a set of single antennas. The output voltages,  $v_{\mathrm{ant}}$, are weighted and summed up to produce the station voltage output as
\begin{gather}
	v_{\mathrm{sta}}(\nu)=\sum_i w_i v_{{\mathrm{ant}}, i}(\nu),
\end{gather}
where
\begin{gather}
	v_{{\mathrm{ant}}, i}(\nu)=\int g(\mathbf{n}, \nu) E(\mathbf{n}, \nu) e^{\im 2\pi \mathbf{n}\cdot \mathbf{x}_i/\lambda} d\mathbf{n}\label{eqn_single_ant_voltage}
\end{gather}
is the complex voltage output of $i$-th antenna in the station at position $\mathbf{x}_i$, $g(\mathbf{n}, \nu)$ is the voltage response to the quasi-monochromatic incident plane wave from direction $\mathbf{n}$ within a narrow band centered on frequency $\nu$, $E$ is the scalar electric field strength at the origin, $\lambda=c/\nu$ is the corresponding wavelength, and $\im^2=-1$.
The single antennas in the station are assumed to be identical, just like the treatments in many other works \citep[e.g., ][]{2015ApJ...815...88S, 2015ApJ...809...18P, 2020MNRAS.499...52M}.
As polarization is not concerned with global CD/EoR signal detection, only scalar theory is considered here.
When necessary, one can easily modify the following deductions for polarization measurements.

The basic idea of measuring the global CD/EoR signal is to acquire the voltage signal output by a single antenna,
and if $\partial g(\mathbf{n}, \nu)/\partial \nu=0$, then the power spectral density of the antenna output (i.e., the antenna temperature spectrum) is proportional to the sky-averaged radio spectrum, which is modeled to be the sum of the foreground component, the noise component, and the global CD/EoR signal.
Theoretically, the foreground component can be easily subtracted by performing a power-law or low-degree log-space polynomial model fitting \citep[e.g., ][]{2006ApJ...650..529W, 2013ApJ...763...90W}.
However, when the equipment effects involve, the frequency-dependent antenna gain will be entangled in the antenna temperature spectrum and prevent the above measurements from being performed.
By tuning the weights of every single antenna in the station of an aperture array, one can make the station beam pattern approximately frequency independent within a given frequency range so that the antenna temperature spectrum can precisely reflect the sky-averaged radio spectrum.
The detailed principle is deduced as follows.

The auto-correlation $S(\nu)$ of the station output voltage signal is calculated as
\begin{gather}
	S(\nu)\equiv \left < v_{\mathrm{sta}}(\nu)v^*_{\mathrm{sat}}(\nu)\right >\\
	=\left <\sum_i w_i v_{{\mathrm{ant}}, i}(\nu) \sum_j w^*_j v^*_{{\mathrm{ant}}, j}(\nu)\right >\label{eqn_dbf_paradigm}\\
	=\sum_{i,j} w_i w^*_j \left <v_{{\mathrm{ant}}, i}(\nu) v^*_{{\mathrm{ant}}, j}(\nu)\right >.\label{eqn_correlation_paradigm}
\end{gather}
By substituting Equation \ref{eqn_single_ant_voltage} into the above equation, we obtain
\begin{gather}
	S(\nu)=\sum_{i,j} w_i w^*_j \left <\int g(\mathbf{m}, \nu) E(\mathbf{m}, \nu) e^{\im 2\pi \mathbf{m}\cdot \mathbf{x}_i/\lambda} d\mathbf{m} \right .\notag \\\left .\times \int g^*(\mathbf{n}, \nu) E^*(\mathbf{n}, \nu) e^{-\im 2\pi \mathbf{n}\cdot \mathbf{x}_j/\lambda} d\mathbf{n} \right >\\
	=\sum_{i,j} w_i w^*_j \int |g(\mathbf{n}, \nu)|^2\left <|E(\mathbf{n}, \nu)|^2\right >e^{\im 2\pi \mathbf{n}\cdot (\mathbf{x}_i-\mathbf{x}_j)/\lambda} d\mathbf{n}\\
	\allowdisplaybreaks[1]
	=\int \left [|g(\mathbf{n}, \nu)|^2 \sum_{i,j} w_i w^*_j e^{\im 2\pi \mathbf{n}\cdot (\mathbf{x}_i-\mathbf{x}_j)/\lambda} \right ]\left <|E(\mathbf{n}, \nu)|^2\right >d\mathbf{n}\\
	\allowdisplaybreaks[1]
	=\int \left [|g(\mathbf{n}, \nu)|^2 \left |\sum_{i} w_i e^{\im 2\pi\mathbf{n}\cdot \mathbf{x}_i/\lambda }\right |^2\right ]\left <|E(\mathbf{n}, \nu)|^2\right >d\mathbf{n}.
\end{gather}

The (unnormalized) station beam pattern can be defined as
\begin{gather}
	P_{\mathrm{sta}}(\mathbf{n}, \nu)\equiv |g(\mathbf{n}, \nu)|^2 \left |\sum_{i} w_i e^{\im 2\pi\mathbf{n}\cdot \mathbf{x}_i/\lambda }\right |^2\\
	\equiv|g(\mathbf{n}, \nu)|^2 \left |G_a(\mathbf{n}, \nu)\right |^2.
\end{gather}
Considering that $T_{\mathrm{sky}}(\mathbf{n}, \nu)\propto <|E(\mathbf{n}, \nu)|^2>$, we have
\begin{gather}
	T_{{\mathrm{sat}}}(\nu)=\frac{\int P_{\mathrm{sat}}(\mathbf{n}, \nu)T_{\mathrm{sky}}(\mathbf{n}, \nu)d\mathbf{n}}{\int P_{\mathrm{sat}}(\mathbf{n}, \nu)d\mathbf{n}}.\label{eqn_T_sta}
\end{gather}

With a proper array configuration and a set of finely tuned $w_i$'s, one can make
\begin{gather}
	\frac{\partial P_{\mathrm{sat}}(\mathbf{n}, \nu)}{\partial \nu}\approx 0,
\end{gather}
which will be described in detail later in Section \ref{sec_example}.

We assume that the antennas are placed on a regular $s\times s$ grid as
\begin{gather}
	\mathbf{x}_{p,q}=d(p \mathbf{e}_x+q \mathbf{e}_y), \label{eqn_ant_xy}
\end{gather}
where $s$ is chosen to be an odd number, the unit vector $\mathbf{e}_x$ and $\mathbf{e}_y$ point to the east and the north direction, respectively, and $d$ is the antenna spacing.
Then we obtain
\begin{gather}
	G_a(\mathbf{n}, \nu)=\sum_{p,q}w_{p,q} \exp[\im2\pi d_\lambda(p n_x+q n_y)],\label{eqn_array_gain_grid}
\end{gather}
where $d_\lambda\equiv d/\lambda$, $n_x\equiv\mathbf{n}\cdot \mathbf{e}_x$, and $n_y\equiv\mathbf{n}\cdot \mathbf{e}_y$.
Apparently $G_a(\mathbf{n}, \nu)$ can be represented as the discrete Fourier transform (DFT) to the complex weights on a regular grid.
If $d_\lambda > 0.5$, grating lobes caused by the aliasing effect may appear \citep[e.g.,][]{mailloux2017phased}.

Theoretically, given a desired final station beam pattern $\hat{P}_{\mathrm{sat}}(\mathbf{n}, \nu)$, one can obtain $w_{p,q}$'s through an inverse DFT (IDFT).

\subsection{Determining Antenna Weights}
\label{ssec_ant_weights}
The weights of all the antennas are determined according to the desired station beam pattern $\hat{P}_{\mathrm{sta}}$.
In the following sections, we set a more strict constraint to the weights as all $w$'s are real numbers and
\begin{gather}
	w_{p,q}=w_{-p, q}=w_{p, -q},\label{eqn_w_symmetry}
\end{gather}
so that Equation \ref{eqn_array_gain_grid} can be rewritten as
\begin{gather}
	G_a(\mathbf{n}, \nu)=\sum_{0\le p,0 \le q} w_{p,q} [(2-\delta_{p,0}) \cos(2\pi d_\lambda p n_x) \notag\\\times (2-\delta_{q,0})\cos(2\pi d_\lambda q n_y)].\label{eqn_array_gain_grid_real}
\end{gather}
Without loss of generality, $w_{p,q}$'s are normalized by $w_{0,0}$, i.e., the weight of the central antenna, so that $w_{0,0}\equiv 1$.
Although values of $w$'s can be determined by performing an IDFT, incomplete aperture plane coverage and numerical errors lead the total station beam pattern deviates from the desired one.
Practically $w_{p, q}$'s are determined by solving an optimization problem
\begin{gather}
	\mathbf{w}=\arg \min_{\mathbf{w}} \int\left |P_{\mathrm{sat}}(\mathbf{n}, \nu; \mathbf{w})-\hat{P}_{\mathrm{sat}}(\mathbf{n}, \nu)\right |^2 d\mathbf{n}.\label{eqn_fobj}
\end{gather}
This optimization problem is solved for each frequency channel independently.
Other objective functions may also be used, but we do not perform the test here for conciseness.

As this is a high-dimensional optimization problem, the dimension of which is $(s+1)^2/4-1$, we choose to use the particle swarm optimization \citep[PSO,][]{10.1162/EVCO_r_00180} algorithm.
The IDFT result is used as an initial guess of the PSO algorithm.

\section{From DBF to Short Spacing Interferometer}
\label{sec_dbf2corr}
\subsection{Forming the Beam with an Equivalent Interferometer}
\label{ssec_equivalent_interferometer}
By comparing Equations \ref{eqn_dbf_paradigm} and \ref{eqn_correlation_paradigm}, one may notice that the measurement of the station auto-correlation signal can be performed with two methods  \citep[e.g.,][]{ruigrok2017cross}: summing the weighted single antenna voltage signal and then calculating the auto-correlation of the station output, namely
\begin{gather} \label{method1}
	S(\nu)=\left <\sum_i w_i v_{{\mathrm{ant}}, i}(\nu) \sum_j w^*_j v^*_{{\mathrm{ant}}, j}(\nu)\right >,
\end{gather}
or summing the weighted auto- and cross-correlations of all antenna pairs, namely
\begin{gather}
	S(\nu)=\sum_{i,j} w_i w^*_j \left <v_{{\mathrm{ant}}, i}(\nu) v^*_{{\mathrm{ant}}, j}(\nu)\right >\label{eqn_correlator_output}.
\end{gather}

The data acquisition of the two methods corresponds to the conventional digital beamforming system and interferometer system, respectively.
We name the first method the DBF paradigm (Equation \ref{method1}) and name the second method the correlation paradigm (Equation \ref{eqn_correlator_output}).
Although these two paradigms are equivalent to each other mathematically, the correlation paradigm has an advantage over the DBF paradigm: the weights of the antennas can be determined and applied offline.
It gives a second chance to process the data if the instrumental calibration improves after the observation is performed.
The correlation paradigm also has a disadvantage: the original time resolution has been lost since the signals acquired have already been integrated,  which is fortunately not required in global CD/EoR signal detection.

In the correlation paradigm, the correlation result is only related to the relative position between the two antennas involved. Thus the number of data acquisition channels can be reduced (to save costs) by canceling redundant baselines. However, one antenna might be in more than one baseline and can be removed only when other non-redundant baselines are unaffected. All the antennas are checked one by one and removed if allowed; until no more antennas can be removed while keeping the baseline coverage unchanged. Note that the final array configuration is affected by the order of the checking and removing procedure, so it is not uniquely determined.

\subsection{Statistic Error}
According to Equations \ref{eqn_worst_delta_rel} and \ref{eqn_eta_max} in the Appendix, the relative error in the worst case of the correlation paradigm is calculated as
\begin{gather}
	\Delta_{\mathrm{rel}}[S(\nu)]_{\mathrm{max}}\notag\\
	=\frac{1}{\sqrt{\Delta\nu\tau}}\sqrt{\frac{1}{N}+\frac{\sum_{k\neq 0}[\frac{1}{r(i,j)}(\sum_{(i,j)\in b(k)} w_i w_j)^2]}{(\sum_{i} w_i^2)^2 }}\\
	\equiv\frac{\eta_{\mathrm{max}}}{\sqrt{\Delta\nu\tau}}.
\end{gather}
In the next section, we will show that for a set of practical weights, the $\eta_{\max}$ is $\sim 1$ so that the measurement is close to the condition of a single antenna, and the required integration time is around several days.

\begin{figure*}
	\begin{center}
		\includegraphics[width=0.75\columnwidth]{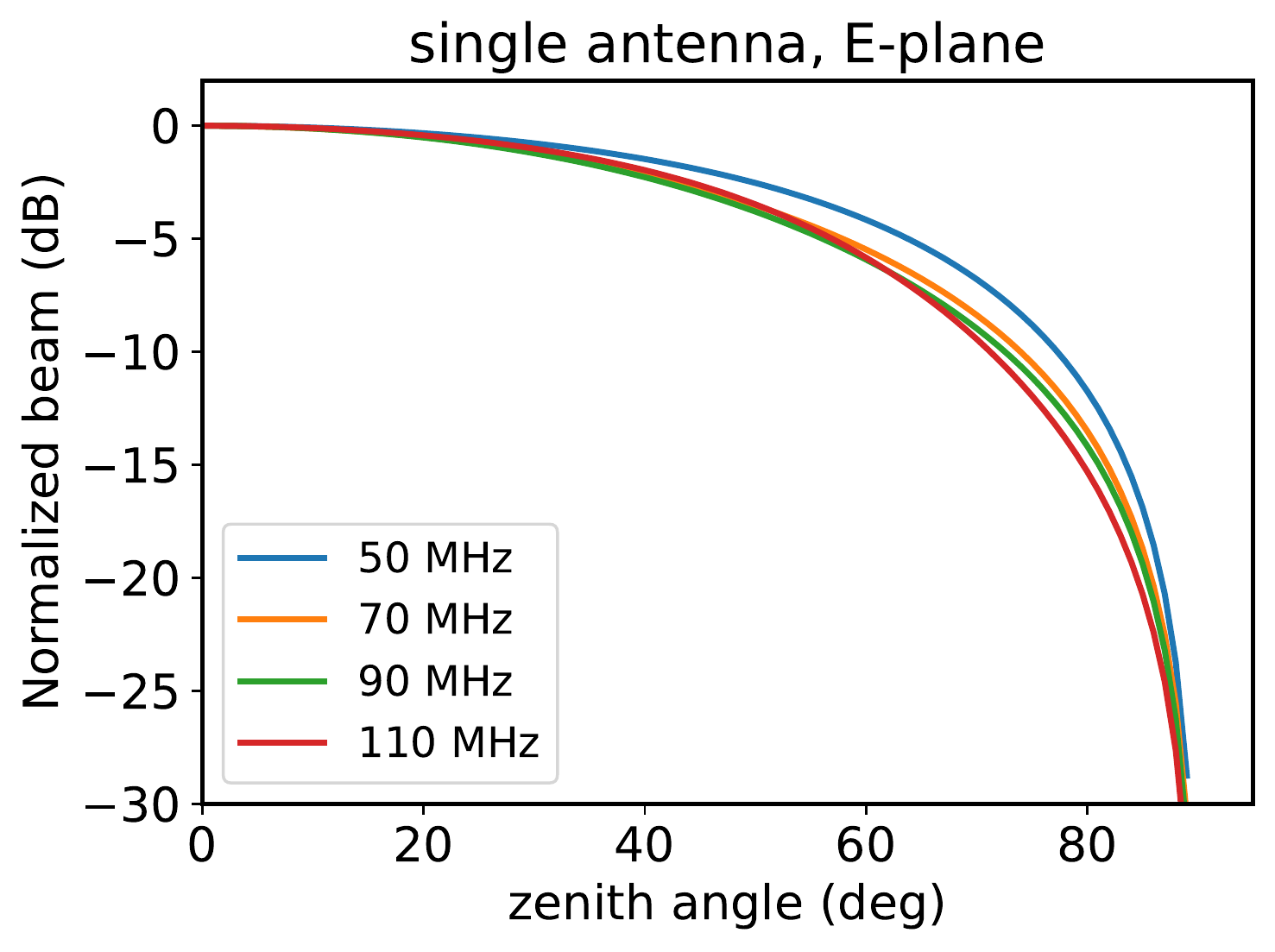}
		\includegraphics[width=0.75\columnwidth]{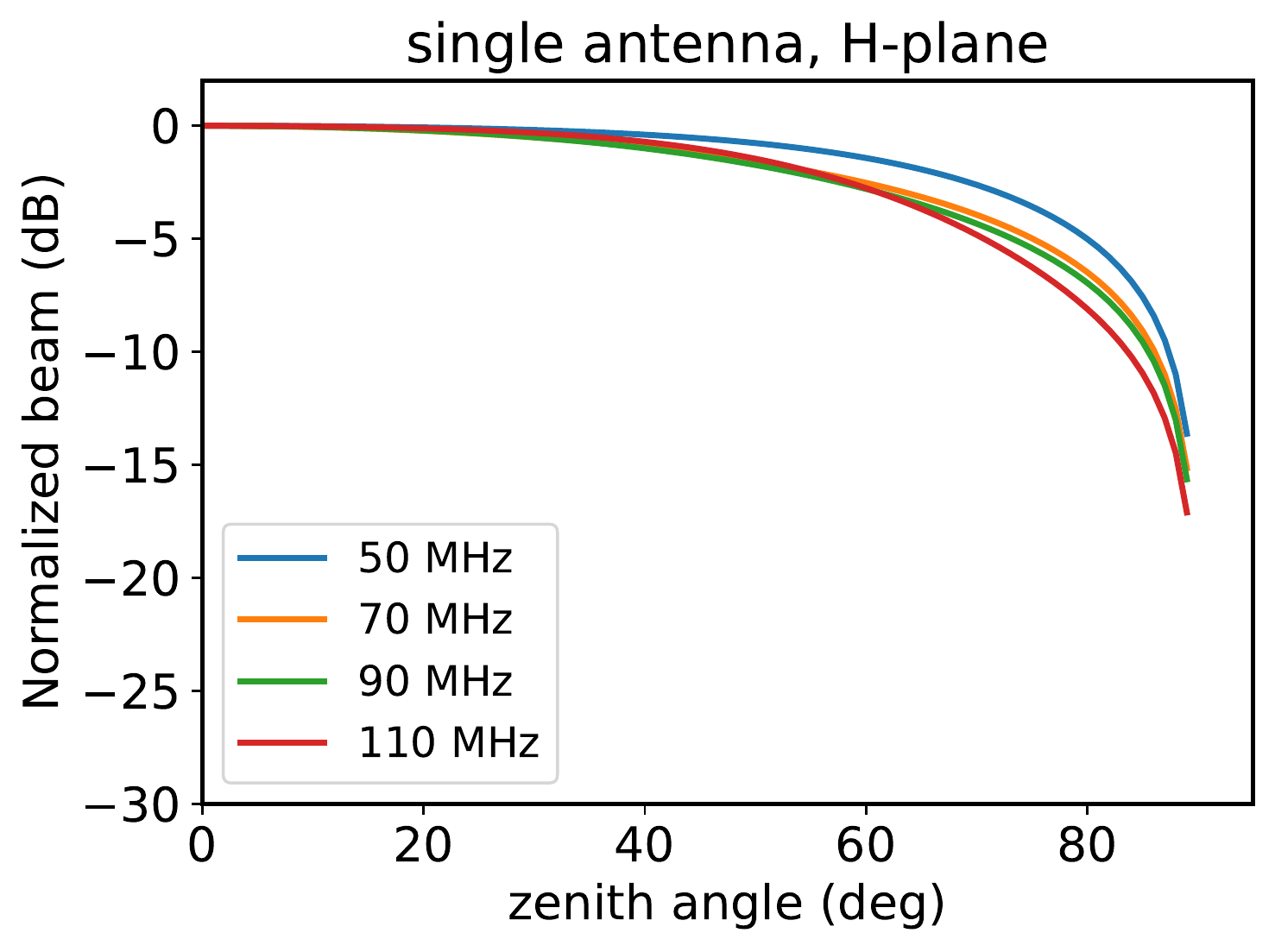}
	\end{center}
	\caption{\label{fig_ant_beam} The profiles of the normalized single antenna beam pattern in the E-plane and the H-plane.}
\end{figure*}

\begin{figure}
	\begin{center}
		\includegraphics[width=1.\columnwidth]{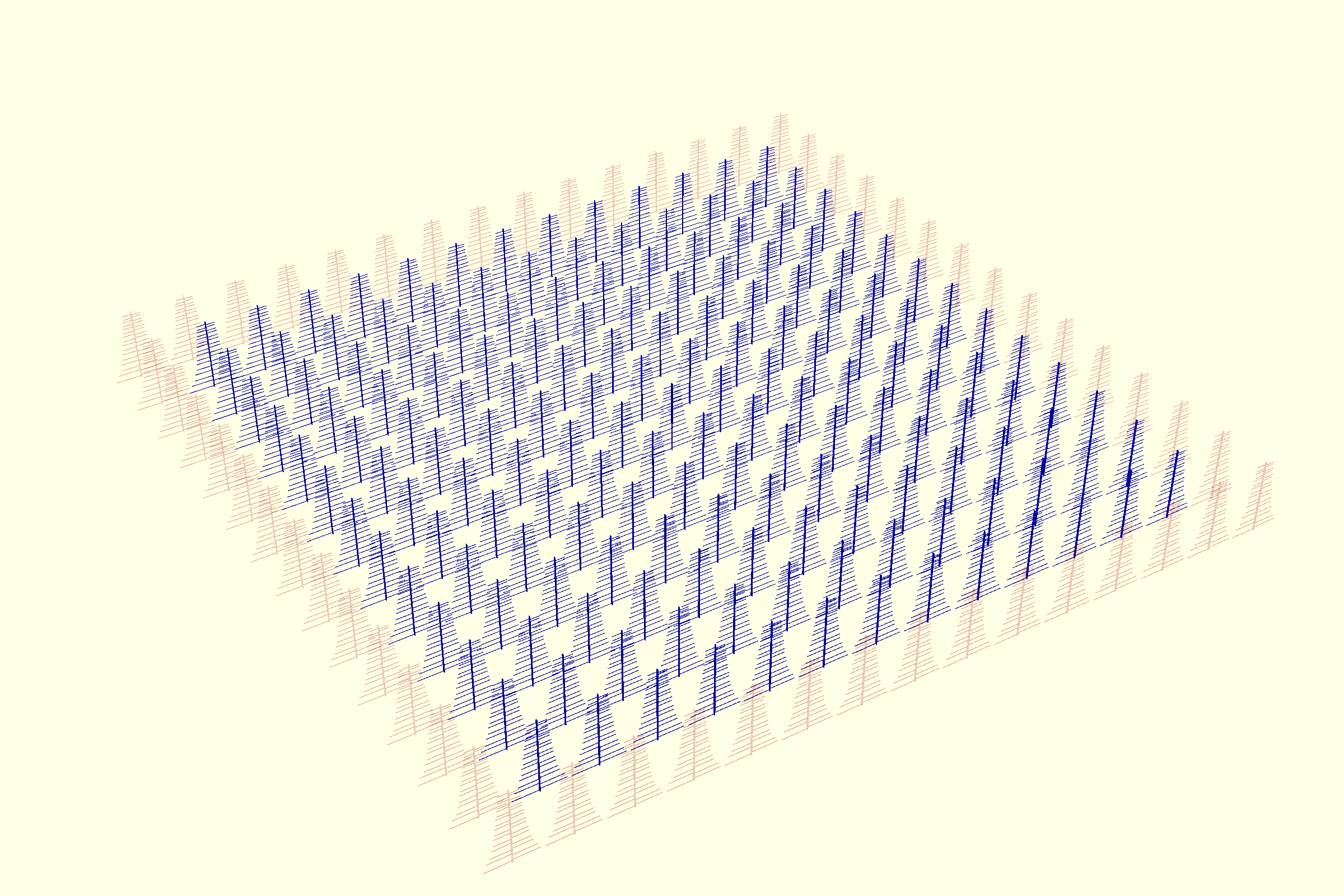}
	\end{center}
	\caption{\label{fig_full_array} The configuration of the full DBF array. The signal of the antennas in blue is measured, while the antennas in red are only connected to terminal resistors.}
\end{figure}

\section{An Example}
\label{sec_example}
In this section, we present an example design of the DBF array, optimize the weights, and cut down the redundant baselines to reduce the number of required data acquisition channels.
We assume that some facilities of the 21 Centimetre Array (21CMA;  \citealt{2010ApJ...723..620W, 2016RAA....16...36H, 2016ApJ...832..190Z}) can be directly used in this experiment, so the single antenna design and the array configuration take this into account.

The example shown here is to demonstrate the feasibility of forming frequency-independent beam patterns with an array, not necessarily an optimal design.
Both the single antenna design and array configuration can be optimized further.
Besides, engineering challenges are not thoroughly handled in this paper, which will be briefly discussed in Section \ref{ssec_engineering_challenges}.

\subsection{The Design of Single Antennas in the Array}
\label{ssec_example_single_ant}
21CMA is an interferometer working in the $50-200$ MHz band.
Each antenna station of 21CMA is composed of 127 log-periodic antennas optimized for the working band.
It is equipped with a digital correlator with 40 data acquisition channels.

We choose to use a modified version of the log-periodic antenna used in the 21CMA array in this example.
The length of the longest dipole of the log-periodic antenna is shortened to fulfill the mechanical constraint of the array configuration.
The axes of the single antennas are steered to the zenith rather than their original pointings in 21CMA, i.e., the north celestial pole.
The beam patterns of the single antenna are calculated with {\sc nec++}, a C++ implementation of the \textit{Numerical Electromagnetics Code}\footnote{\url{http://elec.otago.ac.nz/w/index.php/Necpp}}(NEC).
We show the beam patterns of 50, 70, 90, and 110 MHz in Figure \ref{fig_ant_beam}.
The NEC code is implemented based on the method of moments \citep[e.g.,][]{davidson2010computational}.
It is worth evaluating if results with higher accuracy can be obtained with other choices, such as the finite element method (FEM) and the finite difference time domain (FDTD).
Though log-periodic antennas are wide-band antennas, the single antenna beam pattern in this example is still significantly frequency-dependent.
In Section \ref{ssec_evaluating_example}, we will show that the artificial spectral structure, which is caused by the frequency-dependent beam pattern of a single antenna, may exceed the expected amplitude of the global CD/EoR signal.

\subsection{Array Configuration in the DBF Paradigm}
\label{ssec_dbf_array_cfg}
The configuration of the DBF array is set to be a $13\times 13$ regular grid, i.e., $s=13$, with the spacing to be 1.5 m.
In Section \ref{ssec_removing_redundant_baselines}, we will show that $s=13$ ensures a total of 37 data acquisition channels are required after removing all possible redundant baselines; it can fit well into the current 40-channel correlator of the 21CMA.
The antennas' response on the square grid's boundary may be different from that of the antennas inside, so a set of antennas that are only connected to matching terminal resistors can be placed right outside the grid (Fig. \ref{fig_full_array}).

According to Section \ref{ssec_equivalent_interferometer}, the DBF array can be equivalent to an interferometer so that the redundant baselines can be canceled.
We show the array configuration after removing all possible redundant baselines later (\S \ref{ssec_removing_redundant_baselines}).

\begin{figure*}
	\begin{center}
		\includegraphics[width=0.9\textwidth]{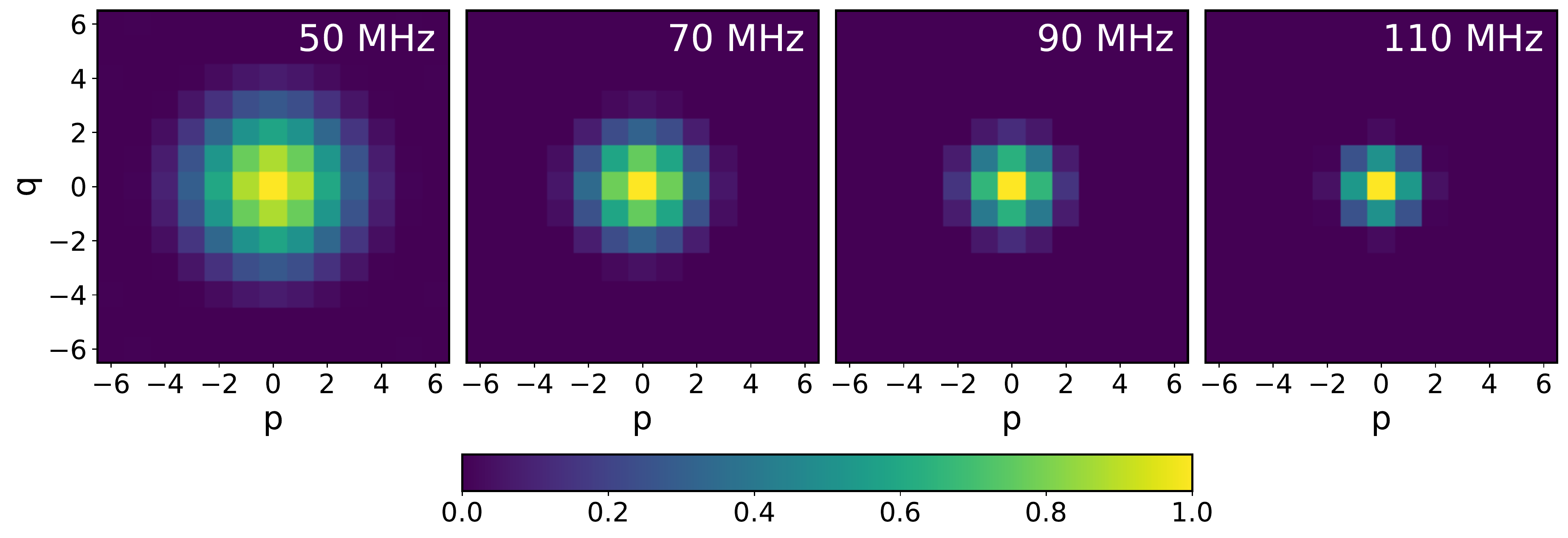}
	\end{center}

	\caption{\label{fig_weights}The weights of four frequency channels, the four pictures are all of the size $13\times 13$ pixel$^2$, corresponding to the antennas on the $13\times13$ regular grid. The value of each pixel equals the weight of each corresponding antenna.}
\end{figure*}

\begin{figure*}
	\begin{center}
		\includegraphics[width=0.75\columnwidth]{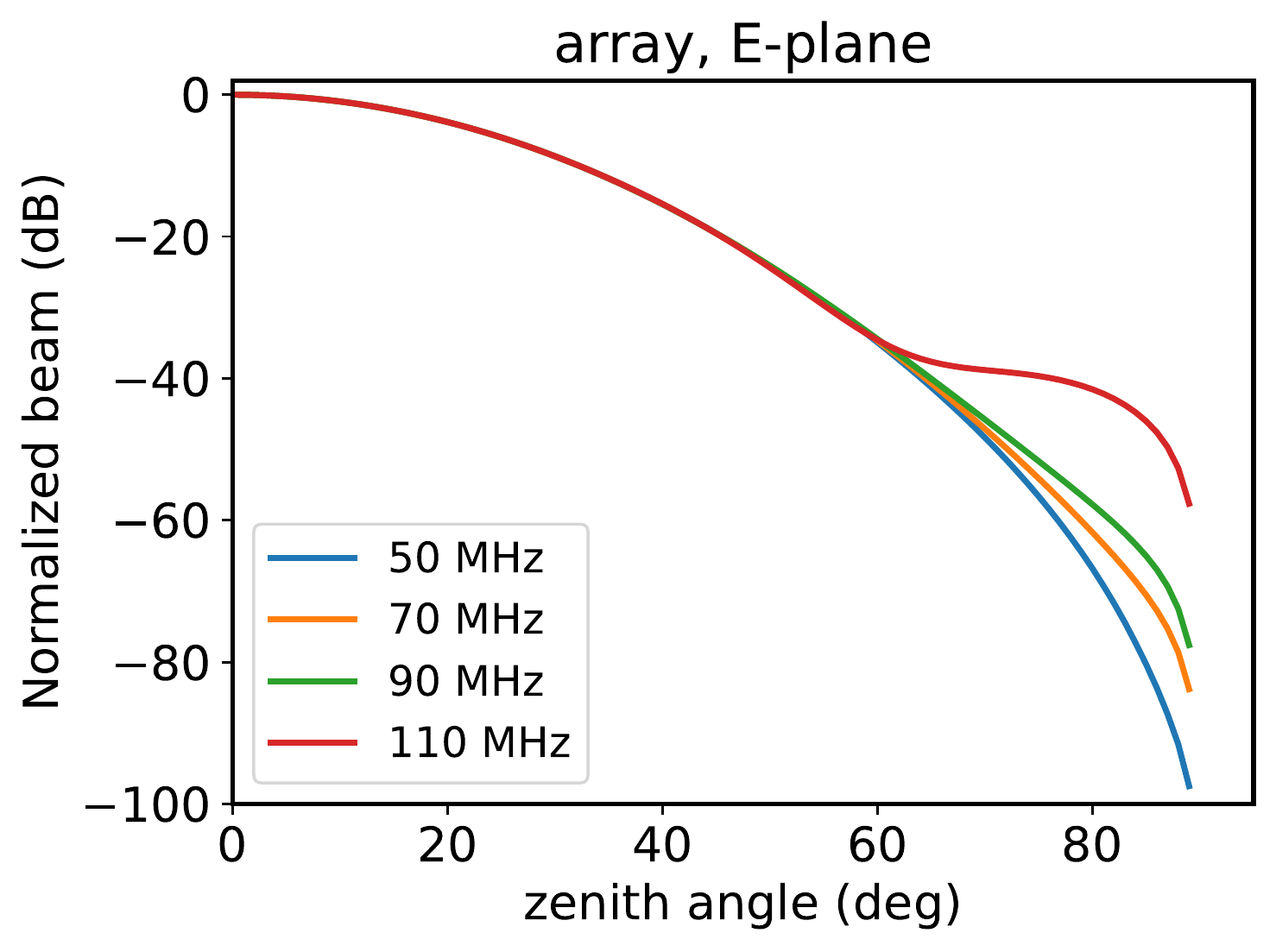}
		\includegraphics[width=0.75\columnwidth]{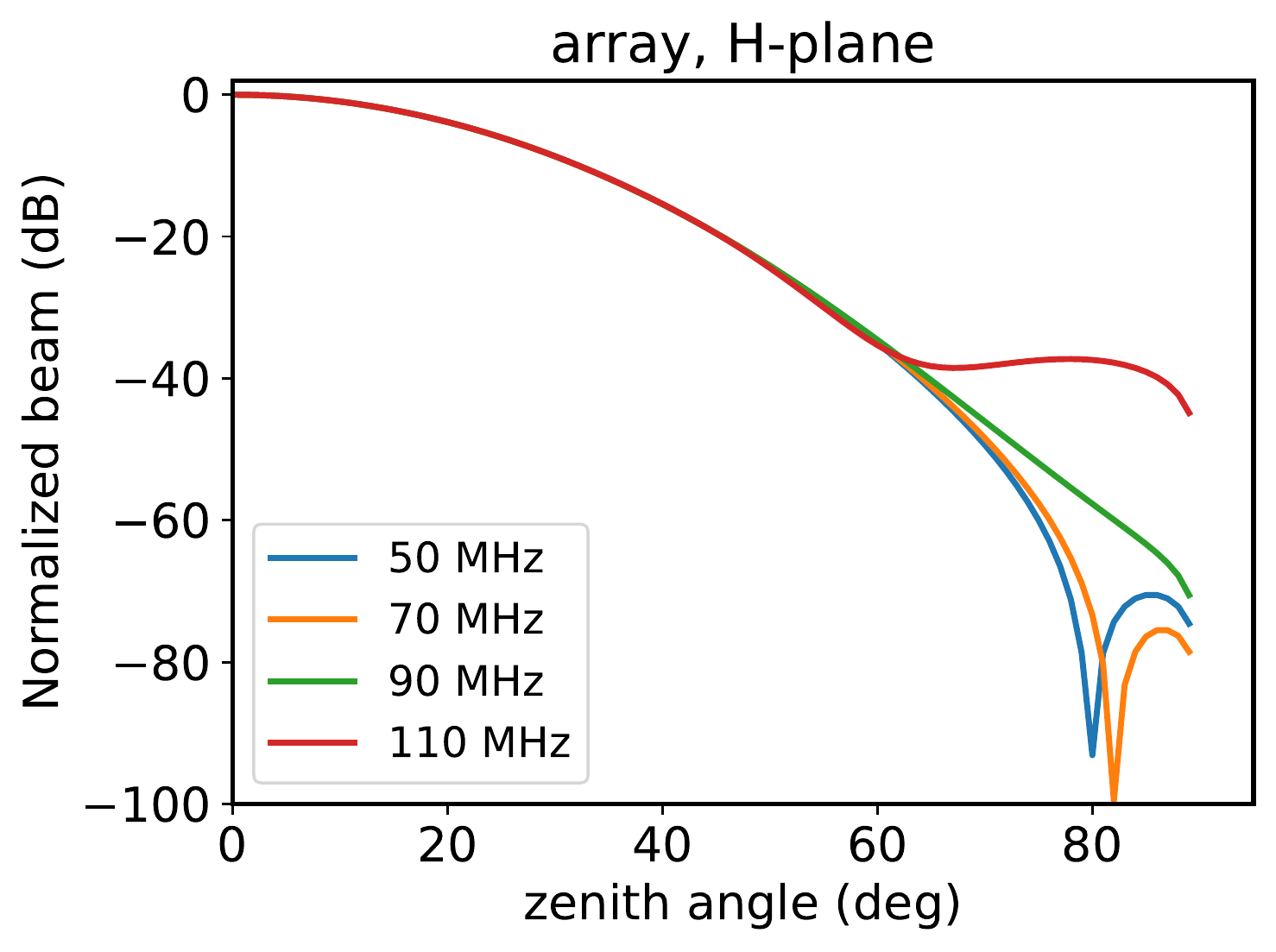}
	\end{center}
	\caption{\label{fig_array_beam} The normalized array beam pattern profiles in the E-plane and the H-plane.}
\end{figure*}

\subsection{Calculating the Weights}
\label{ssec_example_wgt}
The weights are determined by solving the optimization problem Equation \ref{eqn_fobj}.
The code was mainly run on the ARM64 architecture cluster of the China SKA Regional Centre prototype \citep[][]{an2022status}.
About 500 core-hours were consumed.
Note that a full array configuration composed of $s^2$ antennas is assumed here.

We set the desired station beam pattern to be a Gaussian beam as
\begin{gather}
	\hat{P}_{\mathrm{sta}}(\theta)\propto \exp\left(-\frac{\theta^2}{2\theta^2_b}\right),\label{eqn_target_beam}
\end{gather}
where $\theta_b=15^\circ$, and $\theta$ is the zenith angle.
By minimizing the objective function of Equation \ref{eqn_fobj} with the PSO algorithm, we calculate the weight of each antenna in each frequency channel between $50$ and $110$ MHz, with the channel spacing of $1$ MHz,  Figure \ref{fig_weights} shows the obtained weights of $50$ MHz, $70$ MHz, $90$ MHz, and $110$ MHz, respectively; the corresponding profiles of the array beam patterns in the E-plane and the H-plane are shown in Figure \ref{fig_array_beam}.
We notice that in the high-frequency end (i.e., 110 MHz), sidelobes with level $-40$ dB appears. In section \ref{ssec_evaluating_example}, we will show that detecting the global 21 cm signal is feasible with this level of sidelobes.
For frequencies higher than $110$ MHz, we could not find a set of $w$'s that meet the requirements of global CD/EoR detection.
Hopefully, changing the design of the single antennas or optimizing the array configuration (increasing $s$ and/or decreasing $d$) can extend the frequency range.

\subsection{Evaluating the Influence of the Frequency Dependent Beam}
\label{ssec_evaluating_example}
We use the 408 MHz continuous spectra sky map by \cite{2015MNRAS.451.4311R} \citep[originally created by][]{1981A&A...100..209H} as the sky template and set the spectral index to be $\alpha=-2.7$ to simulate the foreground signal in the interesting frequency range as
\begin{gather}
	T_{\mathrm{sky}}(\mathbf{n},\nu)=T_{\mathrm{408~MHz}}(\mathbf{n})\left(\frac{\nu}{408~\mathrm{MHz}}\right)^\alpha.
\end{gather}
Note that the spectral index may have a variety of different values in difference literature, most of which are in the range between $-2.75$ \citep[e.g., ][]{2022MNRAS.509.4923I} and $-2.5$ \citep[][]{2015ApJ...809...18P,2020MNRAS.499...52M}.

\begin{figure*}
	\begin{center}
		\includegraphics[width=0.75\columnwidth]{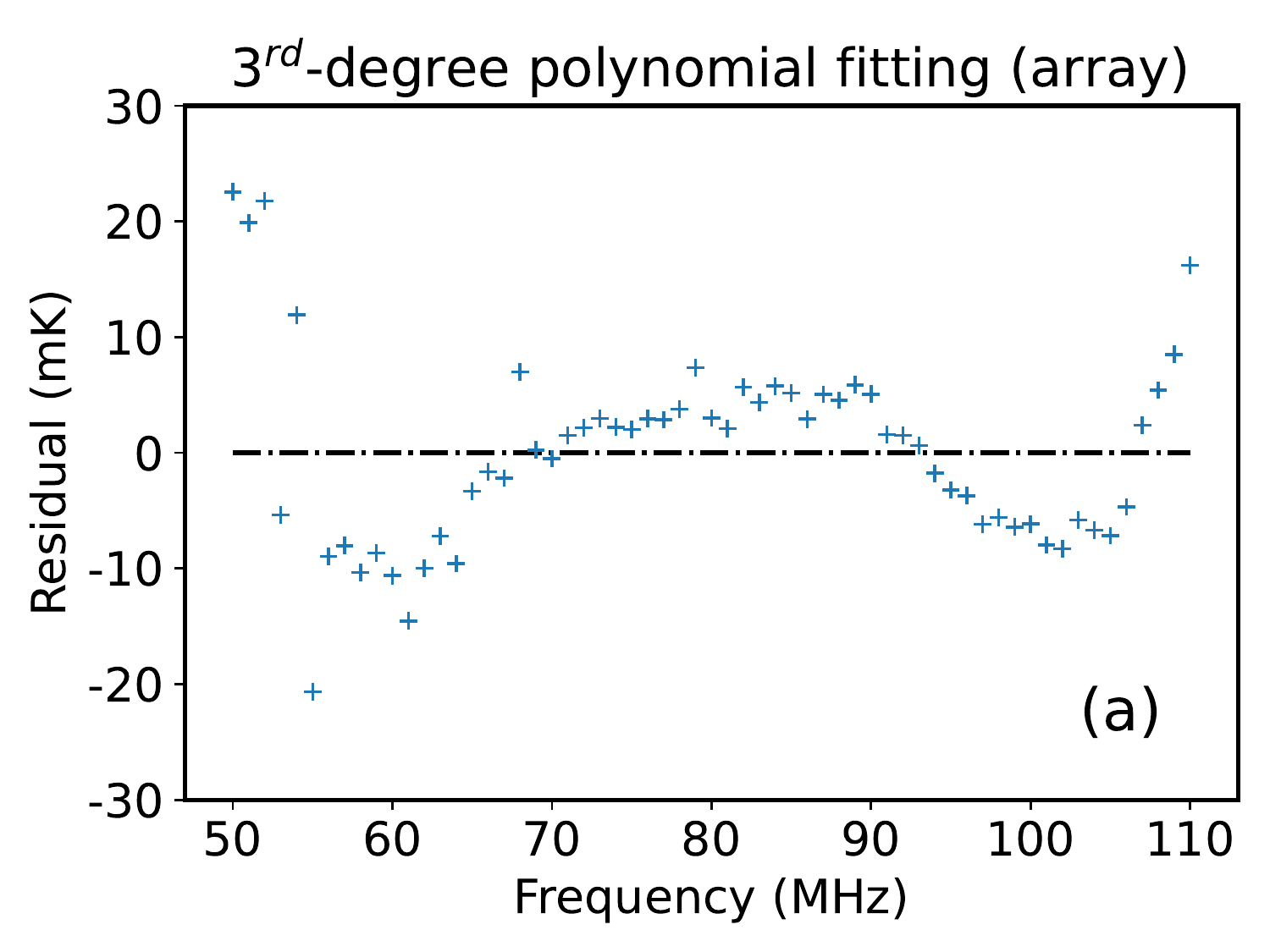}
		\includegraphics[width=0.75\columnwidth]{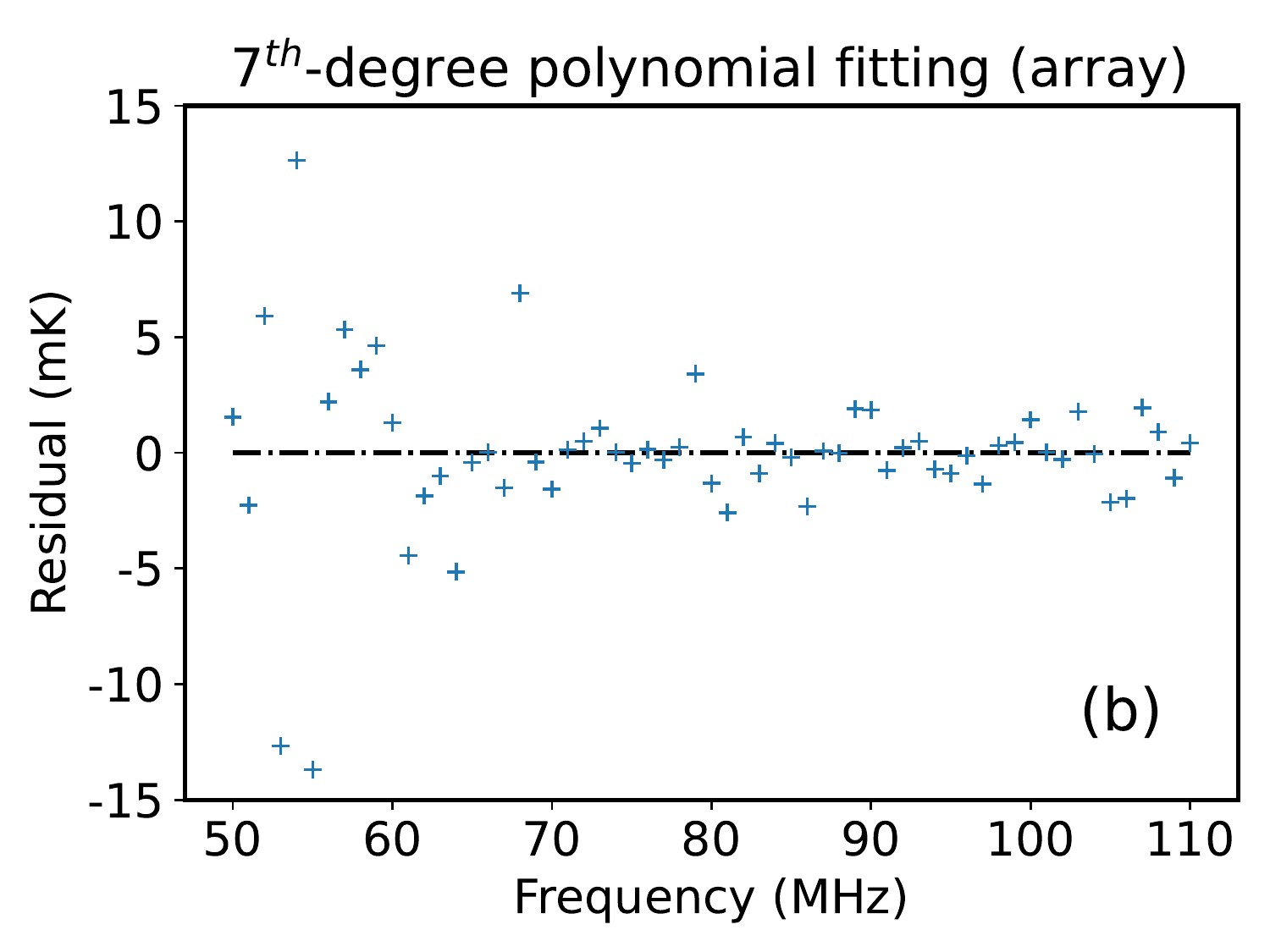}\\
		\includegraphics[width=0.75\columnwidth]{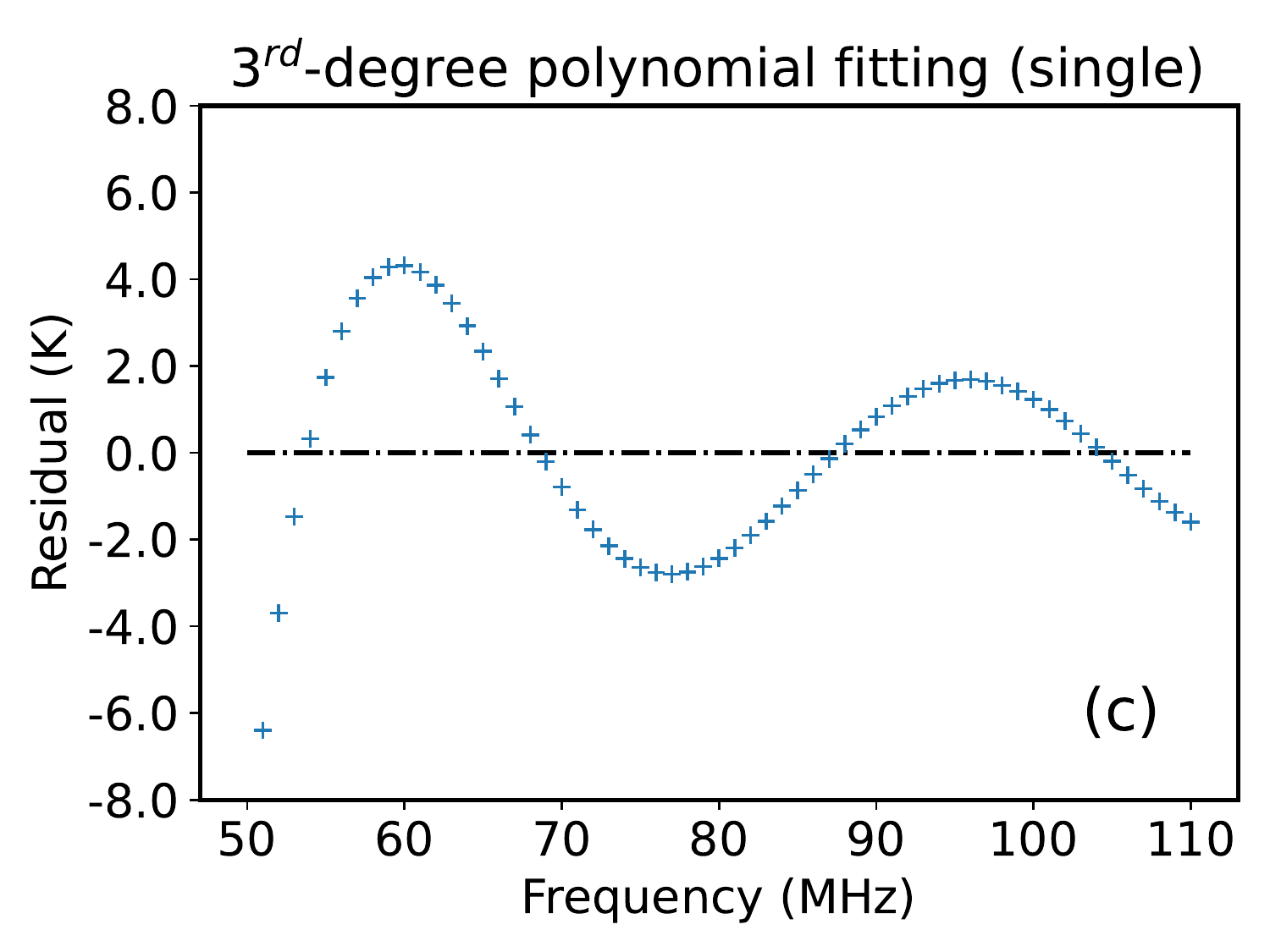}
		\includegraphics[width=0.75\columnwidth]{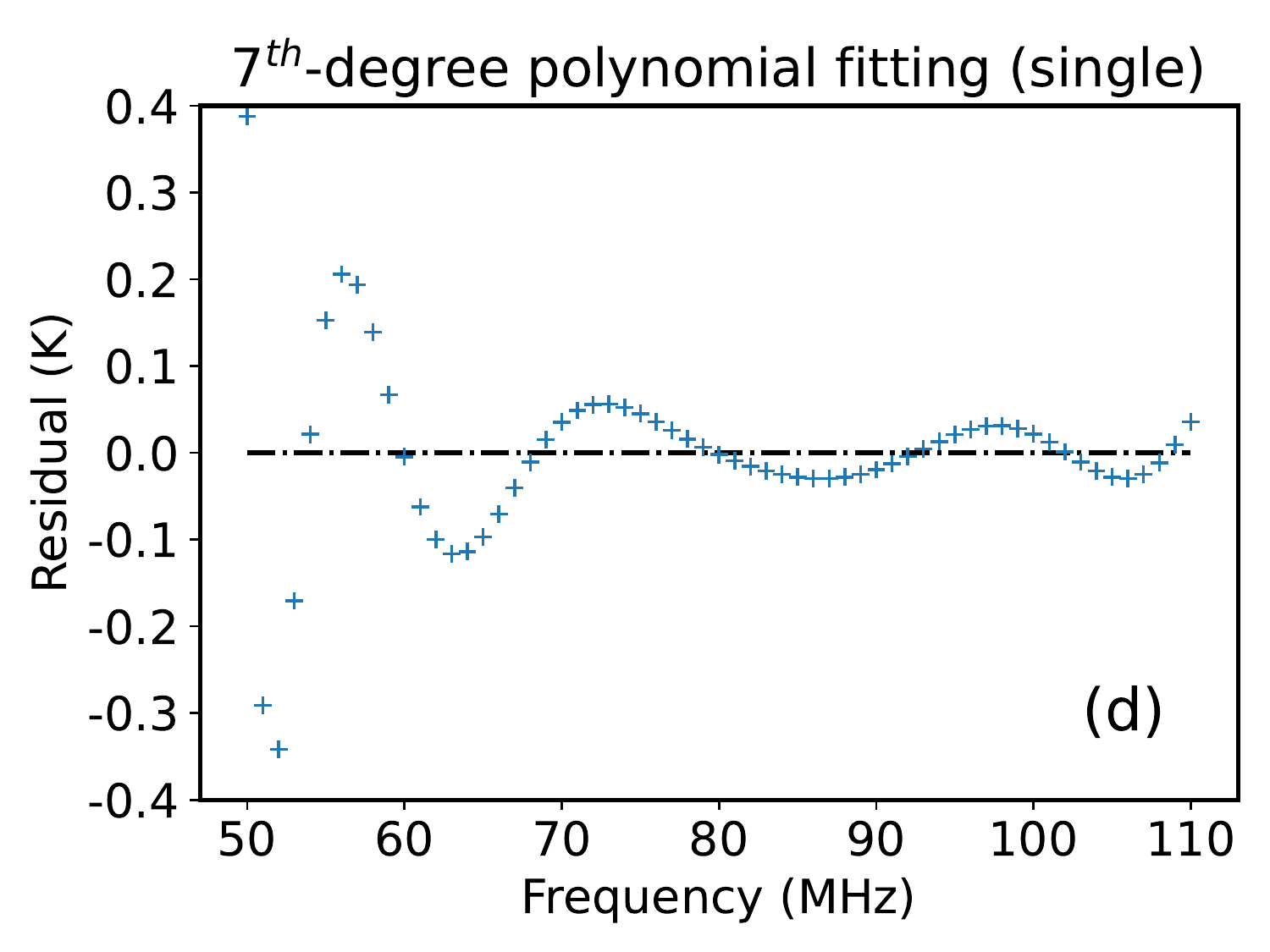}
	\end{center}
	\caption{\label{fig_array_ant_residual} The residual of the $3^{\mathrm{rd}}$-degree (left) and the $7^{\mathrm{th}}$-degree (right) polynomial fitting to the ($\ln(\nu/\mathrm{MHz})$, $\ln (\bar{T}_{\mathrm{sta}}(\nu)/\mathrm{K})$) data set that is simulated based on the array (upper) beam pattern obtained in Section \ref{ssec_example_wgt}, and the single antenna (lower) beam pattern described in Section \ref{ssec_example_single_ant}. Note that the y-axis units are in mK and K for the array and the single antenna data, respectively.}
\end{figure*}

We assume the instrument is placed at a site with a latitude of $45^\circ$N. Then we use Equation \ref{eqn_T_sta} to evaluate the instant station output $T_{\mathrm{sta}}(\nu)$, calculate the 24 hours average spectrum $\bar{T}_{\mathrm{sta}}(\nu)$,  and perform a polynomial fitting to the simulated ($\ln(\nu/\mathrm{MHz})$, $\ln (\bar{T}_{\mathrm{sta}}(\nu)/\mathrm{K})$) data set.
Polynomials with different degrees have been tested in previous works \citep[e.g., ][]{2016MNRAS.461.2847B,2019ApJ...880...26S} to fit the foreground and to mitigate the residual spectral structures caused by the instrumental effects.
Most choices of the degrees of the polynomial are between 3 and 7.
Here we show the residual of a $3^{\mathrm{rd}}$- and a $7^{\mathrm{th}}$-degree polynomial fitting in Figure \ref{fig_array_ant_residual}a and b.
If the global CD/EoR absorption feature (i.e., the dip) is around 100 mK, it should be sufficient to detect this feature.

As a comparison, we use the beam pattern of the single antenna to predict the induced antenna temperature $T_{\mathrm{ant}}$ and perform a $3^{\mathrm{rd}}$- and a $7^{\mathrm{th}}$-degree polynomial fittings, the residual of which are shown in Figure \ref{fig_array_ant_residual}c and d.
The DBF algorithm significantly improves the frequency-independency of the single antennas in the array and makes global CD/EoR signal detection feasible.

\begin{figure}
	\begin{center}
		\includegraphics[width=1.\columnwidth]{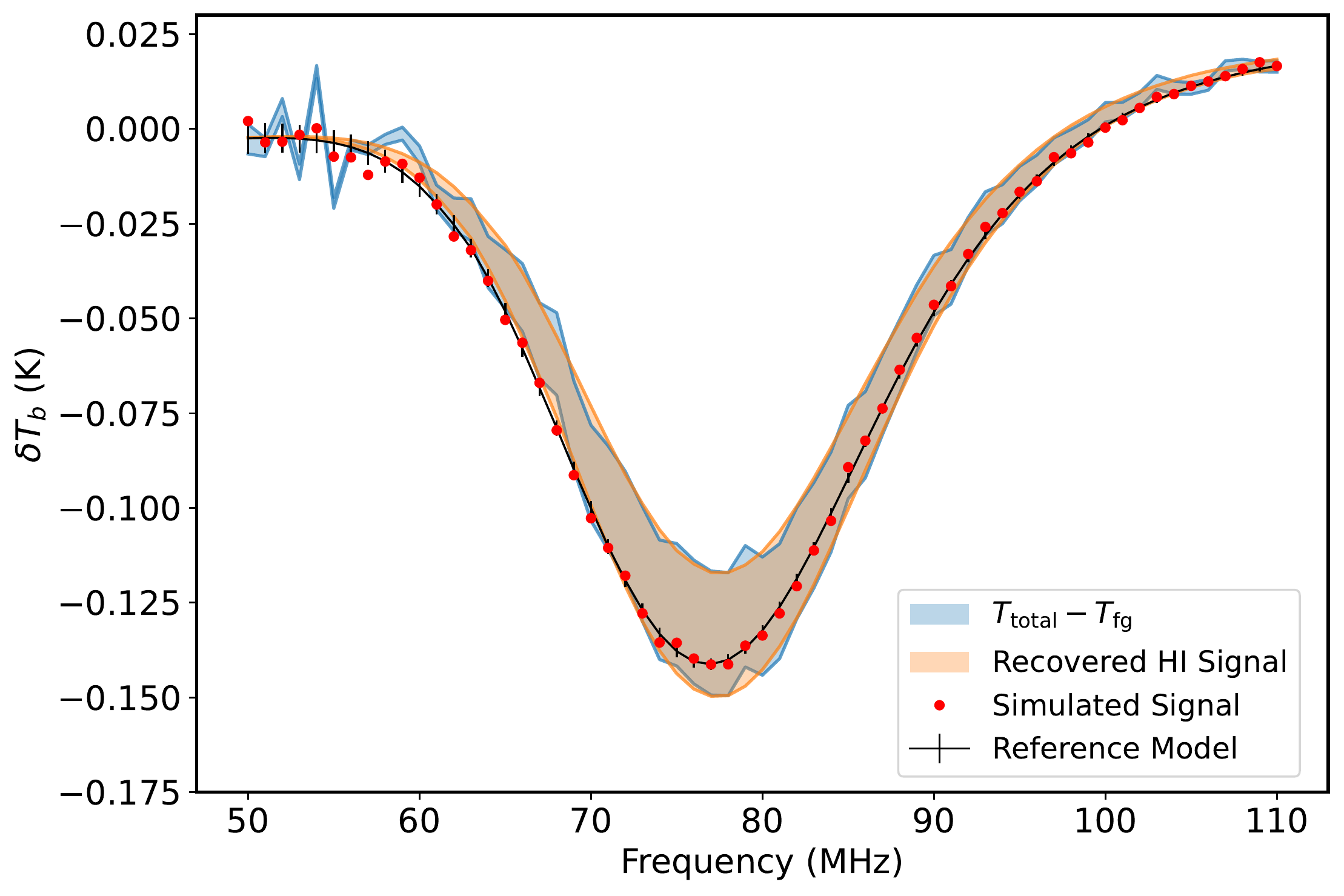}
	\end{center}
	\caption{\label{fig_eor_recovered} A comparison of the input and recovered signal from the simulated data assuming the array to have a configuration as is described in Section \ref{ssec_dbf_array_cfg}.}
\end{figure}

We also test recovering the global CD/EoR signal from the simulated data by using a direct Bayesian inference method described in our previous work \cite{2020MNRAS.492.4080G}.
The global CD/EoR signal model evaluating code was implemented according to the work of \cite{2012ApJ...756...94M} and \cite{2014MNRAS.443.1211M}.
The parameters to simulate the global CD/EoR signal are modified so that the dip falls within the frequency range of $50-110$ MHz.
We have to admit that the actual condition can hardly be so ideal.
We employ the same \textsc{emcee} algorithm \citep{2013PASP..125..306F} as in \cite{2020MNRAS.492.4080G} to perform the Markov-Chain Monte Carlo sampling to recover the signal.
A $7^{\mathrm{th}}$-degree polynomial in log-space is used to model the foreground and any artificial spectral structures introduced by the uncorrected frequency-dependent station beam pattern.
The comparison between the input and the recovered signals is shown in Figure \ref{fig_eor_recovered}.
The global CD/EoR signal is recovered in two ways: (1) by calculating the difference between the total signal $T_{\mathrm{total}}$ and the foreground model $T_{\mathrm{fg}}$, and (2) by directly evaluating the numerical global CD/EoR model.
The characters of the input signal have been well recovered.
Notably, the parameter space may show degeneracy for some other combinations of the global CD/EoR models and foreground models.  More thorough simulation tests may be necessary to ensure the detection of the CD/EoR signal, but beyond the concern of this paper.

\subsection{Removing Redundant Baselines}
\label{ssec_removing_redundant_baselines}
\begin{figure}
	\begin{center}
		\includegraphics[width=1.\columnwidth]{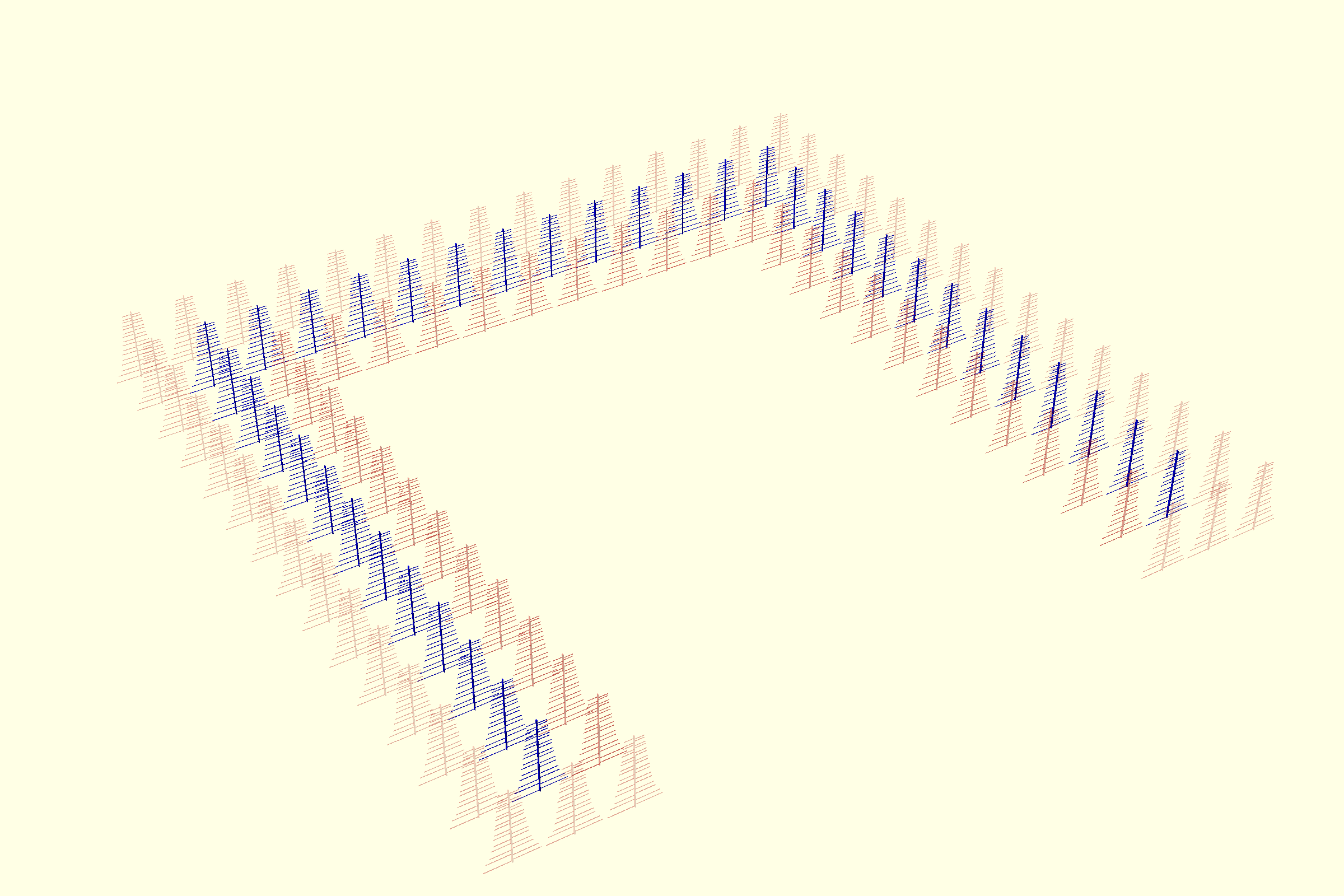}
	\end{center}
	\caption{\label{fig_shrinked_array} The array configuration after removing all possible redundant baselines.}
\end{figure}

According to Section \ref{ssec_equivalent_interferometer}, the DBF array can be equivalent to an interferometer by replacing the DBF data acquisition system with a correlator.
The number of data acquisition channels can be reduced by removing the redundant baselines.
The array configuration after removing all possible redundant baselines is shown in Figure \ref{fig_shrinked_array}.
There are 37 antennas left, so the data acquisition can be performed with the 21CMA correlator directly, which has 40 data acquisition channels.

\subsection{Error Estimation}
The statistic uncertainty is estimated based on the array configuration described in section \ref{ssec_removing_redundant_baselines}, i.e., all possible redundant baselines have been removed.

\begin{figure}
	\begin{center}
		\includegraphics[width=0.75\columnwidth]{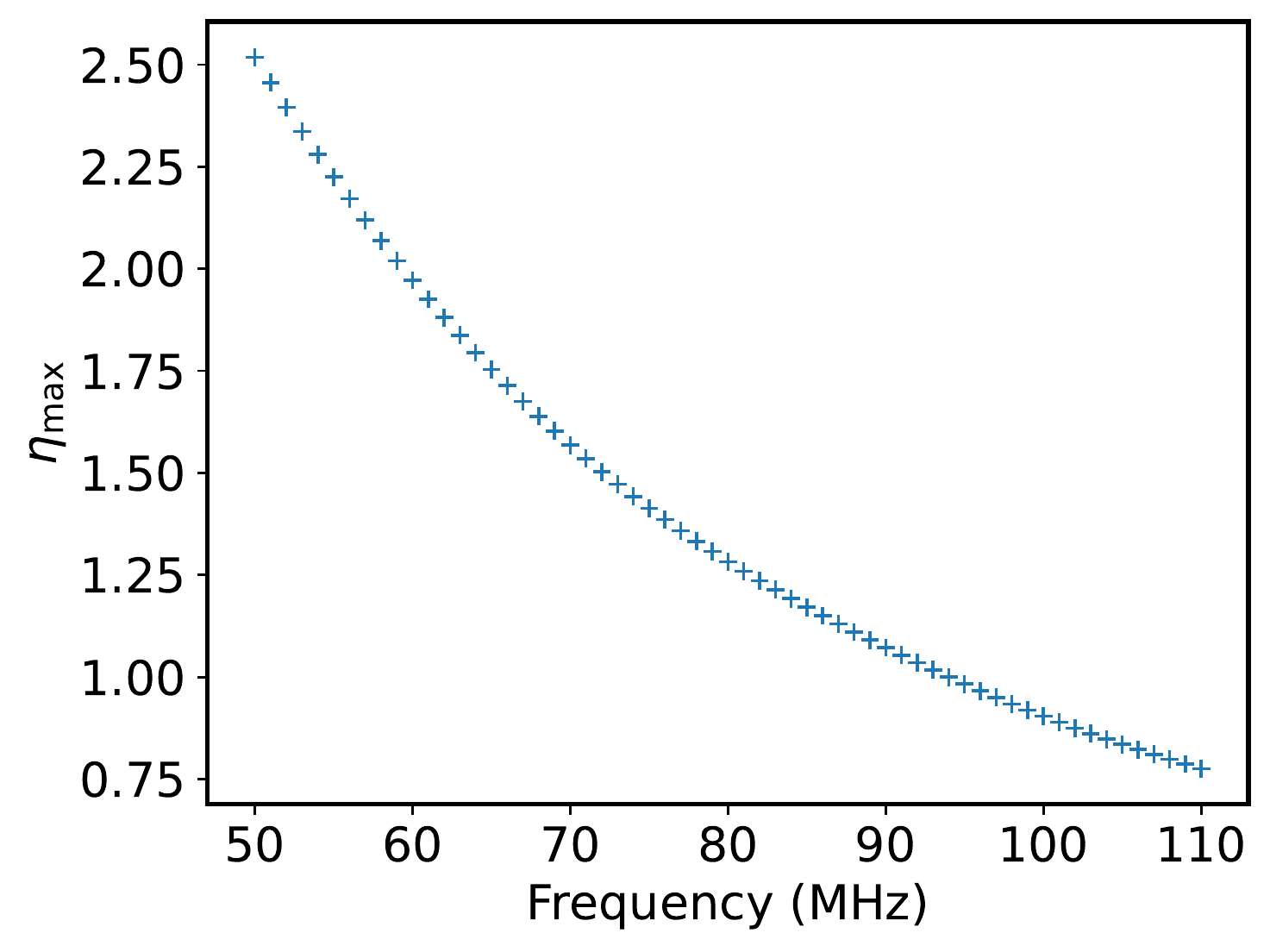}
	\end{center}
	\caption{\label{fig_eta_max} $\eta_{\max}$ calculated for each frequency channel between $50-110$ MHz.}
\end{figure}

We calculate the $\eta_{\max}$ using Equation \ref{eqn_eta_max} and the weights obtained in Section \ref{ssec_example_wgt}. The result is shown in Figure \ref{fig_eta_max}.
The $50$ MHz channel has the largest $\eta_{\max}$, so we use this value to estimate the statistic error for a ten-day observation as an example.
By using Equation \ref{eqn_worst_delta_rel}, and let $\eta_{\max}\approx 2.5$, $\tau=10$ days, $\Delta\nu=1~\mathrm{MHz}$, the relative statistic error of $S(\nu)$ is calculated to be
\begin{gather}
	\Delta_{\mathrm{rel}}[S(\nu)]_{\max}=2.7\times10^{-6}.
\end{gather}
Assuming the absorption feature of the global CD/EoR signal is about $100$ mK, the sky-averaged Galactic foreground is $10^4$ K in the worst case; the required relative statistic error should be less than $10^{-5}$.
So the above array design should be sufficient to constrain the global CD/EoR signal if the instrument calibration error is well controlled.

\section{Discussion}
\label{sec_discussion}
\subsection{Do We Really Need an Array?}
As shown in Section \ref{ssec_evaluating_example}, a well-designed array does help suppress the frequency-dependency of the beam pattern.
But antenna arrays are usually much more expensive and complicated than single antenna systems.

On the single antenna side, the frequency-dependent beam pattern can be suppressed by narrowing down the working frequency range, i.e., dividing the whole frequency range into several sub-bands and building single antennas for each of them. However, if the character bandwidth of the global CD/EoR signal features (e.g., the dip feature) happens to be broader than the sub-bands, the detection can be more difficult.

Therefore, whether to use an array to detect the global CD/EoR signal depends on the compromise among several factors, including the cost, the desired instrument performance, and the technical capacity.

\subsection{An Non-Square Array}
Since the acquired antenna temperature spectra are averaged within a period longer than 24 hours, the station beam pattern profiles in the east-west direction are integrated out eventually, and only the frequency-dependency of the profiles in the north-south direction matters; it is not strictly true due to the array is not located on the equator but is still acceptable in our condition.
Thus it is not necessary to restrict the array configuration to be a full $s\times s$ square array.
The antenna number along the east-west direction can be reduced to $1$ so that we could have a $1\times s$ 1D linear array deployed along the local meridian.
And still, with the procedure described in Section \ref{sec_dbf2corr}, the number of antennas can be reduced while the baseline coverage is kept unchanged.
With this method, the frequency-dependency of the 24-hour averaged station beam patterns can hopefully be better suppressed with the same number of data acquisition channels.

\subsection{Forming Achromatic Beam with SKA LFAA Station?}

The future SKA Low-Frequency Aperture Array (LFAA) antenna station utilizes DBF technology to form the station beam, making it possible to perform global CD/EoR signal detection.
\cite{2022arXiv221004693P} simulated the end-to-end drift-scan observation of the radio sky at 50--100 MHz using a zenith-phased array in the background of the EDA2 array, a precursor instrument for SKA.
They concluded that an EDA2-like array with $\mathcal{O}(10^5)$ antennas can form station beam patterns with side lobe level $<-50$ dB, so that can be used to detect the global CD/EoR signal.
The huge required number of antennas makes it hard to implement.
The simulation result of our example design described in Section \ref{sec_example} suggests that a properly set desired station beam pattern, a carefully chosen array configuration, and a set of optimized antenna weights may enable the detection of global CD/EoR signal with a small number of antennas.

According to the SKA system requirement specification\footnote{\url{https://www.skao.int/sites/default/files/documents/d3-SKA-TEL-SKO-0000008-Rev11_SKA1SystemRequirementSpecification_0.pdf}}, 256 antennas (SKA1-SYS\_REQ 2139) will be randomly (SKA1-SYS\_REQ 3339) distributed within a circular region with a diameter about 35 m (SKA1-SYS\_REQ 2140).
If this configuration is chosen, the average distance between antennas is $\sim 1.94$ m, which is larger than the half wavelength of $78$ MHz.
This could make it hard to form desired beam pattern at frequency $>78$ MHz.
A similar reason may cause the EDA2-like array to need $\mathcal {O} (10^5)$ antennas to detect the global CD/EoR signal.
Nevertheless, it is worth evaluating the performance of the future LFAA station by using the antenna weight optimizing procedure described in Section \ref{ssec_ant_weights}; we will leave this in future work.

\subsection{A Possible Method to Suppress the Grating Lobes}
Limited by the geometric dimension of single antennas, the requirement of $d_\lambda\leq 0.5$ cannot always be met.
In our example array (Section \ref{sec_example}), $d_\lambda$ is limited by the length (123.6 cm) of the longest dipole of the log-periodic antenna.

A possible strategy is rotating the antenna about the vertical axis by $45^{\circ}$. It will allow us to place the single antennas closer to build a denser array; the grating lobes can be further suppressed.
However, this will bring two shortages: (1) when the array becomes denser, the coupling and crosstalk between neighboring antennas will be more severe, and thus it becomes harder to predict the beam pattern of the single antennas, and (2) rotating the antennas by $45^{\circ}$ will destroy the symmetry of the single antenna beam pattern, then the symmetry constraints over the $w_{p,q}$'s described as Equation \ref{eqn_w_symmetry} no longer work if we still require the final array beams to be symmetric (as is described by Equation \ref{eqn_target_beam}). Then the number of free parameters of the optimization problem to solve $w$'s (i.e., Equation \ref{eqn_fobj}) will increase to $s^2-1$ (almost multiplied by $4$), and thus the optimization problem will be harder to solve.

Note that only making the array denser might not improve the overall quality of the formed beams due to the equivalent aperture becoming smaller. A compensatory method is increasing the number of antennas, i.e., enlarging the parameter $s$.
This will require a larger number of data acquisition channels.
This effort requires further evaluation to value its necessity.

\subsection{Suggestions about Antenna Design and Spacing}
The final station beam pattern is comprehensively determined by (1) array configuration, (2) antenna weights $w_{p,q}$, and (3) the beam pattern of single antennas, while most efforts of this work are focused on the first two factors.
The single antennas should have a beam pattern as independent from the frequency as possible.
However, this design target may be limited by other factors.
For example, considering that the antennas are rather close to each other, the dimensions of the single antennas should not be too large to fit into the array, and this factor puts some constraints on the single antenna design.

About the detailed antenna beam pattern shape, as we mentioned in Section \ref{ssec_dbf_principle}, the array should have a $d_\lambda \leq 0.5$; otherwise, grating lobes may appear.
Limited by the dimension of single antennas,  this requirement cannot always be fulfilled, especially at the high-frequency end.
If the single antenna beam pattern can concentrate to the zenith direction, the grating lobes arise by a $d_\lambda>0.5$, which usually have large zenith angles, can be suppressed.

\subsection{Engineering Challenges}
\label{ssec_engineering_challenges}
\subsubsection{An Array in the Real World}
In previous computations and most other works \citep[e.g., ][]{2015ApJ...815...88S, 2015ApJ...809...18P, 2020MNRAS.499...52M, 2020JAI.....950008D}, the single antennas in an array are assumed to be identical,
but in the real world, this assumption does not always hold.
As we have stated in our previous work \cite{2020MNRAS.492.4080G}, any gain calibration error  $>10^{-4}$ may significantly bias the detection result or make the detection completely impossible.
Similarly, any difference between the single antennas in electronic/mechanical aspects might introduce errors comparable to or larger than the target CD/EoR signal.

The above computations assume that the single antenna beam pattern $|g(\mathbf{n}, \nu)|^2$ can be precisely measured beforehand, but for antennas working in the meter-wave band, this kind of measurement is challenging.
Calibrating the beam pattern with a bright radio source does not work in the case of low-frequency single antennas because of their small effect area and wide beam.
Neither is it easy to perform a precise anechoic chamber measurement because (1) the absorption material used to build meter-wave anechoic chamber is usually not good enough to emulate the free space, and (2) ambient facilities (e.g., other antennas, the ground plane) around the antenna being calibrated has to be considered, which is hard to be fit into a common microwave darkroom.
The recent development of drone-based antenna measurement technology \citep[e.g., ][]{paonessa2015uav, 2015PASP..127.1131C} may be used to perform such kind of calibration; however, the accuracy is to be validated.
The tolerance of inaccuracy in the single antenna beam pattern measurement will be discussed in our future work.

For an array composed of a set of closely placed antennas, like the one we described in Section \ref{sec_example}, the coupling or crosstalk between neighboring antennas is an issue that cannot be ignored.
When designing single antennas, this effect should be taken into account.
When performing an actual beam pattern measurement, the single antenna under test should also be placed inside the array.

\subsubsection{The Challenges in the Analog Front-end}
The antenna temperature induced by the foreground emission is four orders higher than that induced by the CD/EoR signal.
This extreme contrast raises a strict requirement for the analog front-end.
The analog front-end is composed of filters and amplifiers, which are expected to be almost identical, stable enough, and well-calibrated (better than the level of $10^{-4}$ dB).

Besides the active analog devices, the co-axis cables connecting the antennas and the data acquisition systems must be cut precisely.
Considering the extreme requirements of the system calibration \citep[e.g.,][]{2020MNRAS.492.4080G}, an error that is only less than the wavelength is far from sufficient.
How sensitive the result is to the cable length error requires further evaluation in future works.

\subsubsection{Self-Generated and External Weak RFIs}
Different from single antenna experiments such as EDGES \citep[][]{2010Natur.468..796B}, BIGHORNS \citep[][]{2015PASA...32....4S}, SCI-HI \citep[][]{2014ApJ...782L...9V}, SARAS \citep[][]{2013ExA....36..319P}, and PRI$^\mathrm{Z}$M \citep[][]{2019JAI.....850004P}, the data acquisition system for an array that is composed of tens of antennas is much more complicated, as the complexity scales with the square of the number of antennas.
The self-generated radio frequency interference (RFI) can be well controlled in single antenna experiments; on the other hand, the high power consumption of the data acquisition system of an array makes it hard to build the shielding system.

Unlike strong radio RFIs, which can be easily identified and excluded, weak RFIs below the detection threshold are hard to handle (e.g., see Section 4.4 of \citealt{2021MNRAS.505.3698W}), especially for the global EoR detection experiments that have given up sky imaging.
Such RFIs might remain in the flagged data and show the appearance of spectral features that are hard to be explained, which may mislead the CD/EoR signal detection. Further discussion about the RFIs is beyond the scope of this paper; we will leave this in future work.

\section{Conclusions}
\label{sec_conclusion}
This work introduces a method to form achromatic beam patterns with a dense regular DBF array to detect the all-sky averaged 21 cm signal from the Cosmic Dawn and Epoch of Reionization.
Section \ref{ssec_ant_weights} gives a procedure to optimize the antenna weights to form the desired station beam pattern.
The equivalence between a DBF array and an interferometer reduces the number of necessary antennas in the DBF array.

An example array design with antennas arranged in a regular $13\times 13$ square grid is given. We calculate the antenna weights to form the desired station beam pattern and evaluate the performance of global CD/EoR signal detection.
A simulation aiming to evaluate the performance of the example array provides a positive indication towards the feasibility of using the dense regular DBF array to detect the global CD/EoR signal. 
The final array configuration after removing all possible redundant baselines is shown in Figure \ref{fig_shrinked_array}.

We also give some general suggestions about the array design and briefly discuss the engineering challenges that must be overcome in practical experiments.

\section*{Acknowledgements}
This work is supported by National SKA Program of China No. 2020SKA0110200 and 2020SKA0110100.
We acknowledges the supports from NSFC under 11973070 and Key Research Program of Frontier Sciences, CAS, Grant No. ZDBS-LY-7013.

This work used resources of China SKA Regional Centre prototype funded by the National Key R\&D Programme of China (2018YFA0404603).

We would also like to thank Marta Spinelli for her comments on the final manuscript.

\section*{Data Availability}
The code and data underlying this article will be shared on reasonable request to the corresponding authors.

\appendix
\section{Statistic Error of the Correlation Paradigm}
\label{sec_appendix}
Based on the Equation \ref{eqn_correlator_output}, given that all $w$'s are real numbers and under the constraint of Equation \ref{eqn_w_symmetry}, the measured value $S(\nu)$ can be rewritten as
\begin{gather}
	S(\nu)=\sum_i w_i^2 <|v_{\mathrm{ant}, i}|^2>+2\sum_{i<j} w_i w_j <\Re(v_{\mathrm{ant}, i} v_{\mathrm{ant}, j}^*)>\\
	=\sum_i w_i^2 <\Re(v_{\mathrm{ant}, i})^2>\notag\\
	+ \sum_i w_i^2 <\Im(v_{\mathrm{ant}, i})^2>\notag\\
	+2\sum_{i<j} w_i w_j <\Re(v_{\mathrm{ant}, i})\Re(v_{\mathrm{ant}, j})>\notag\\
	+2\sum_{i<j} w_i w_j <\Im(v_{\mathrm{ant}, i})\Im(v_{\mathrm{ant}, j})>\\
	=\sum_{i,j} w_i w_j (<\Re(v_{\mathrm{ant}, i})\Re(v_{\mathrm{ant}, j})>+<\Im(v_{\mathrm{ant}, i})\Im(v_{\mathrm{ant}, j})>),
\end{gather}
where $\Re(\cdot)$ and $\Im(\cdot)$ denote the real and imaginary parts of a complex number, respectively.
Note that a certain pair of antennas say ($i_1$, $j_1$) could have been removed so that $<\Re(v_{\mathrm{ant}, i_1})\Re(v_{\mathrm{ant}, j_1})>$ and $<\Im(v_{\mathrm{ant}, i_1})\Im(v_{\mathrm{ant}, j_1})>$ could be missing.
However, the removing antenna strategy described in Section \ref{ssec_equivalent_interferometer} ensures that there is always at least one equivalent baseline remaining in the array.
No matter whether one baseline is missing, its correlation value is always replaced by the average values obtained from the still remaining equivalent baselines as
\begin{gather}
	\Re(v_{\mathrm{ant}, i_1})\Re(v_{\mathrm{ant}, j_1})=C^r(k)\equiv\frac{1}{r(k)}\sum_{(i,j)\in b(k)} \Re(v_{\mathrm{ant}, i})\Re(v_{\mathrm{ant}, j})\\
	\Im(v_{\mathrm{ant}, i_1})\Im(v_{\mathrm{ant}, j_1})=C^i(k)\equiv\frac{1}{r(k)}\sum_{(i,j)\in b(k)} \Im(v_{\mathrm{ant}, i})\Im(v_{\mathrm{ant}, j}),
\end{gather}
where $k$ is the index of the baseline from the $i_1$-th antenna to the $j_1$-th antenna, $b(k)$ is a function mapping the $k$-th baseline to the set of still existing corresponding antenna pairs $(i,j)$ in the regular grid, $r(k)$ is the number of members in the set $b(k)$. Correspondingly we also define a function $B(k)$, mapping $k$-th baseline to the set of all possible antenna pairs $(i,j)$, whether they are removed or not.

Then
\begin{gather}
	S(\nu)=\sum_{k}[(<C^r(k)>+<C^i(k)>)\sum_{(i,j)\in B(k)} w_i w_j] .
\end{gather}

In a certain frequency channel with a bandwidth of $\Delta\nu$, centered on frequency $\nu$, the variance of the real (and also imaginary) part of the complex voltage follows a normal distribution with the variance of
\begin{gather}
	\Var[\Re (v_{ant, i})]=\frac{1}{2}R k_{\mathrm{B}}T_{\mathrm{ant}}\Delta \nu,
\end{gather}
and the expected value of $0$, $R$ is the system impedance, $k_{\mathrm{B}}$ is the Boltzmann constant, and $T_{\mathrm{ant}}$ is the antenna temperature.
In the meter-wave band, the system temperature is dominated by the Milky Way foreground, so we do not distinguish between the antenna and system temperatures.

Given that $\Re(v_{\mathrm{ant}, i})$'s, $\Im(v_{\mathrm{ant}, i})$'s, $\Re(v_{\mathrm{ant}, j})$'s, and $\Im(v_{\mathrm{ant}, j})$'s follow identical normal distributions and $\rho_{i,j}$ is the correlation coefficient between $\Re(v_{\mathrm{ant}, i})$ and $\Re(v_{\mathrm{ant}, j})$ (and also between $\Im(v_{\mathrm{ant}, i})$ and $\Im(v_{\mathrm{ant}, j})$), $|\rho_{i,j}|_{i\neq j}\le 1$, and $\rho_{i,i}=1$.
So the variance
\begin{gather}
	\Var[\Re(v_{\mathrm{ant}, i})\Re(v_{\mathrm{ant}, j})]=\frac{1+\rho^2_{i,j}}{4}(R k_\mathrm{B}T_{\mathrm{ant}}\Delta \nu)^2,
\end{gather}
and
\begin{gather}
	\Var[\Im(v_{\mathrm{ant}, i})\Im(v_{\mathrm{ant}, j})]=\Var[\Re(v_{\mathrm{ant}, i})\Re(v_{\mathrm{ant}, j})].
\end{gather}

Then
\begin{gather}
	\Var[C^r(k)]=\Var[C^i(k)]=\frac{1+\rho^2_{i,j}}{4r(k)}(R k_\mathrm{B}T_{\mathrm{ant}}\Delta \nu)^2
\end{gather}

During a period of integration time $\tau$, $C^r(k)$ and $C^i(k)$ are measured $\Delta\nu\tau$ times respectively.
Thus
\begin{gather}
	\Var[<C^r(k)>+<C^i(k)>]=\frac{1+\rho^2_{i,j}}{2r(k)\Delta\nu\tau}(R k_\mathrm{B}T_{\mathrm{ant}}\Delta \nu)^2
\end{gather}
\begin{gather}
	\Var[S(\nu)]=\frac{(R k_\mathrm{B}T_{\mathrm{ant}}\Delta \nu)^2 }{\Delta\nu\tau}\sum_{k}[\frac{1+\rho^2_{i,j}}{2r(k)}(\sum_{(i,j)\in B(k)} w_i w_j)^2]\notag\\
	=\frac{(R k_\mathrm{B}T_{\mathrm{ant}}\Delta \nu)^2 }{\Delta\nu\tau}\{ \frac{1}{N}(\sum_i w_i^2)^2\notag\\
    +\sum_{k\neq 0}[\frac{1+\rho^2_{i,j}}{2r(k)}(\sum_{(i,j)\in B(k)} w_i w_j)^2]\},
\end{gather}
and the standard deviation of $S(\nu)$ is
\begin{gather}
	\sigma[S(\nu)]=\frac{R k_{\mathrm{B}}T_{\mathrm{ant}}\Delta\nu}{\sqrt{\Delta\nu\tau}}\notag\\\times\sqrt{\frac{1}{N}(\sum_i w_i^2)^2+\sum_{k\neq 0}[\frac{1+\rho^2_{i,j}}{2r(k)}(\sum_{(i,j)\in B(k)} w_i w_j)^2]}.
\end{gather}

The expected value of $S(\nu)$ is dominated by auto-correlations so that it can be approximated as
\begin{gather}
	E[S(\nu)]\approx\sum_{i,j} w_i w_j \delta_{i,j} (<\Re(v_{\mathrm{ant}, i})\Re(v_{\mathrm{ant}, j})>\notag\\+<\Im(v_{\mathrm{ant}, i})\Im(v_{\mathrm{ant}, j})>)\\
	=\sum_{i} w_i^2 (<\Re(v_{\mathrm{ant}, i})^2>+<\Im(v_{\mathrm{ant}, i})^2>)\\
	=Rk_{\mathrm{B}} T_{\mathrm{ant}}\Delta\nu\sum_{i} w_i^2.
\end{gather}

Then we can calculate the relative uncertainty as
\begin{gather}
	\Delta_{\mathrm{rel}}[S(\nu)]\equiv\frac{\sigma[S(\nu)]}{E[S(\nu)]}\\
	=\frac{1}{\sqrt{\Delta\nu\tau}}\sqrt{\frac{1}{N}+\frac{\sum_{k\neq 0}[\frac{1+\rho^2_{i,j}}{2r(i,j)}(\sum_{(i,j)\in B(k)} w_i w_j)^2]}{(\sum_{i} w_i^2)^2 }},
\end{gather}
in the worst case
\begin{gather}
	\Delta_{\mathrm{rel}}[S(\nu)]_{\max}
	=\frac{1}{\sqrt{\Delta\nu\tau}}\sqrt{\frac{1}{N}+\frac{\sum_{k\neq 0}[\frac{1}{r(i,j)}(\sum_{(i,j)\in B(k)} w_i w_j)^2]}{(\sum_{i} w_i^2)^2 }}. \label{eqn_worst_delta_rel}
\end{gather}

We define
\begin{gather}
	\eta_{\mathrm{max}}\equiv\sqrt{\frac{1}{N}+\frac{\sum_{k\neq 0}[\frac{1}{r(i,j)}(\sum_{(i,j)\in B(k)} w_i w_j)^2]}{(\sum_{i} w_i^2)^2 }},\label{eqn_eta_max}
\end{gather}
to represent the ratio of the relative uncertainty of the array output to that of a single antenna in the worst case.

\bibliographystyle{mnras}
\bibliography{ms} 





\bsp	
\label{lastpage}
\end{document}